\documentclass[a4paper, 12pt]{article}

\usepackage[top=2cm, bottom=2cm, left=2.5cm, right=2.5cm]{geometry}
\usepackage{amssymb}
\usepackage{amsmath}
\usepackage{amssymb,amsfonts,amsthm}
\usepackage{bm}
\usepackage{hyperref}
\usepackage{setspace}
\usepackage[justification=centering]{caption}
\usepackage{float}
\usepackage{xcolor}
\usepackage{changes}
\usepackage{multirow}
\usepackage{longtable}
\usepackage{supertabular} 
\usepackage{dcolumn}
\usepackage{pdflscape}
\usepackage{float}
\usepackage[shortlabels]{enumitem}
\usepackage{longtable}
\usepackage{tabularx,calc}
\usepackage{booktabs}
\usepackage{acro}
\usepackage{setspace}
\usepackage{afterpage}
\usepackage{lscape}
\usepackage{mathtools}
\usepackage{pst-plot, pstricks} 
\usepackage{fancybox,amssymb,color}
\usepackage{subfigure}
\usepackage{framed}

\usepackage{titling}
\usepackage[bottom]{footmisc} 
\usepackage[]{authblk}
\usepackage[title]{appendix}

\usepackage[backend=biber, style=apa, maxcitenames=2, sorting=nyt, sortcites=false]{biblatex}
\usepackage[most]{tcolorbox}

\definecolor{Diceblue}{RGB}{52,116,181}

\hypersetup{
	colorlinks = true,
	linkcolor=Diceblue,
	citecolor=Diceblue,
	urlcolor=Diceblue
}

\begin{document}

	
\setlength{\abovedisplayskip}{3pt}
\setlength{\belowdisplayskip}{12pt}
	
	
\setlength{\jot}{0.5cm}

\begin{titlingpage}

\title{Dynamic Consumer Demand at Large Scale}

\author{
Daniel Brunner$^\dagger$,
Florian Heiss$^\ddagger$ and
Anna B. Schmidt$^\ast$
}

\date{\small\today}

\maketitle

\begin{abstract}
		
We study consumer demand in large-scale retail settings with many products, multiple categories and repeated purchase behavior. While inertia and brand loyalty are well documented, existing discrete choice models typically focus on single categories or become computationally infeasible in high-dimensional environments. We propose a dynamic product-level factor model that captures heterogeneity in baseline preferences, price sensitivity and inertia through a shared latent factor structure. By factorizing individual-product coefficients, the model pools information across individuals and categories and allows for correlated heterogeneity. We estimate the model using Bayesian variational inference, enabling scalable estimation with tens of thousands of parameters. In a simulation study calibrated to realistic retail data, we show that the dynamic factor model substantially improves predictive performance relative to static factor models and mixed logit benchmarks, particularly when individual purchase histories are sparse. Accounting for inertia also leads to more elastic demand estimates, underscoring the importance of dynamics for measuring consumer responsiveness. Our results highlight dynamic factor models as a scalable and flexible approach for demand estimation in modern, high-dimensional retail markets.

		

\vspace{0.4cm}
		
 \vspace{0.1cm}\newline

\end{abstract}

\vfill
\noindent\hrulefill \\
{\footnotesize
We gratefully acknowledge financial support by the German Research Foundation (DFG, grant HE 6902/2-1). Computational infrastructure and support were provided by the Center for Information and Media Technology (ZIM) at Heinrich Heine University Düsseldorf. 
\begin{description}
    \setlength{\itemsep}{0pt}
    \setlength{\parskip}{0pt}
    \setlength{\parsep}{0pt}
    \item[$^\dagger$] Heinrich Heine University Düsseldorf, Chair of Statistics and Econometrics, Universitätsstr. 1, 40225 Düsseldorf, Germany; e-mail: \texttt{\href{mailto:daniel.brunner@hhu.de}{daniel.brunner@hhu.de}}
    \item[$^\ddagger$] Heinrich Heine University Düsseldorf, Chair of Statistics and Econometrics, Universitätsstr. 1, 40225 Düsseldorf, Germany; e-mail: \texttt{\href{mailto:florian.heiss@hhu.de}{florian.heiss@hhu.de}}
    \item[$^\ast$] Heinrich Heine University Düsseldorf, Chair of Statistics and Econometrics, Universitätsstr. 1, 40225 Düsseldorf, Germany; e-mail: \texttt{\href{mailto:anna.brigitte.schmidt@hhu.de}{anna.brigitte.schmidt@hhu.de}}
\end{description}
}

\end{titlingpage}


\onehalfspacing

\section{Introduction}\label{s:intro}
Classical discrete choice models in economics typically assume that consumers evaluate all available alternatives on each decision occasion and select the option that maximizes contemporaneous utility \parencite{train2009discrete}. This framework underlies much of modern demand estimation and has been highly successful in settings where choices are infrequent or where switching costs are explicit. However, a large empirical literature documents systematic deviations from this benchmark, particularly in environments characterized by repeated decisions \parencite{guadagni1983logit, keane1997heterogeneity, dube2010state, bronnenberg2012evolution}. 
In many such settings, consumers display strong inertia and persistence, repeatedly choosing the same option even when switching would be optimal. These patterns raise fundamental questions about how consumer preferences evolve over time and how demand systems should be modeled to capture realistic behavior. 

Empirical evidence for inertia and habit formation has been documented across a wide range of markets. In long-term contract settings such as health insurance and electricity, consumers often remain with incumbent options despite sizable gains from switching, indicating substantial state dependence in choice behavior \parencite{hortaccsu2017power,heiss2021}. Related evidence from consumer packaged goods markets shows that brand loyalty can persist for years and even survive major geographic moves, suggesting that habits are deeply ingrained rather than purely driven by market frictions \parencite{bronnenberg2012evolution}. While much of this literature emphasizes environments with explicit switching costs, similar mechanisms are likely at work in everyday grocery shopping, where switching costs are smaller, more psychological and harder to observe.

Understanding inertia is important not only for descriptive accuracy, but also for economic counterfactuals and policy analysis. Models that fail to account for state dependence tend to confound habit persistence with low price sensitivity, leading to biased elasticity estimates and misleading predictions. This issue is particularly pronounced in industrial organization and competition policy, where demand elasticities play a central role. For example, \textcite{mackay2024consumer} show that accounting for consumer inertia can substantially alter predicted merger effects in consumer goods markets. More broadly, ignoring dynamics may lead regulators to underestimate firms' ability to exploit consumer inertia through pricing or product design.

At the same time, empirical analysis of consumer demand is increasingly based on large-scale retail data \parencite{dubois2022use}. Scanner records, loyalty card data and online transaction logs provide millions of observations across consumers, products, categories and time periods. While these data offer novel opportunities to study consumer behavior, they also raise severe econometric challenges. Fully flexible discrete choice models that allow for individual-product-specific preferences quickly become infeasible due to the curse of dimensionality. Even moderately sized panels would require estimating millions of parameters, making classical estimation approaches computationally intractable.

Existing solutions typically impose a strong structure to restore tractability. A prominent approach is the mixed logit model, which allows for random coefficients to capture preference heterogeneity while keeping product intercepts fixed across consumers \parencite{berry1995automobile}. Although mixed logit models have been widely applied and offer substantial flexibility relative to simple logit specifications, they remain limited in high-dimensional environments. In particular, they do not naturally scale to settings with many categories and products and they impose restrictive assumptions on the correlation structure of heterogeneity across baseline preferences, price sensitivity and dynamic effects. As a result, much of the empirical demand literature has focused on single-category applications, such as yogurt \parencite{ackerberg2001yogurt,villas2007yogurt, hristakeva2022yogurt}, orange juice \parencite{dube2010state}, or ready-to-eat cereal \parencite{nevo2000mergers, backus2021cereal}, abstracting from the joint nature of real-world shopping behavior. While multi-category application such as \textcite{atalay2023scalable} and \textcite{dopper2025rising} estimate demand across a broad set of product markets, they treat each category as a separate demand system and abstract from cross-category substitution and from pooling preference heterogeneity across categories.

Recent work addresses these limitations by introducing factor structure models that pool information across products and categories. By projecting consumers and products into a shared low-dimensional latent space, factor models dramatically reduce the number of parameters while preserving rich heterogeneity. \textcite{Athey2018ttfm} apply factorization techniques to restaurant choices to capture heterogeneous distance elasticities, while \textcite{Ruiz2020shopper} propose a model to uncover substitution and complementarity patterns in large assortments. Most closely related to our paper, \textcite{Donnelly2021counterfactual} develop a nested factorization model that jointly captures within- and across-category substitution patterns and demonstrates a substantial gain in predictive performance over single-category approaches. However, these papers consider static choice models.

This paper builds on the emerging literature by proposing a dynamic product-level factor model that combines within-category inertia with a shared latent factor representation of heterogeneous preferences.
The low-rank structure reduces effective dimensionality by factorizing consumer-product coefficients into consumer-specific latent factors and product-specific loadings, thus pooling information across products and categories \parencite{Ruiz2020shopper,Donnelly2021counterfactual}.
Bayesian estimation is performed using \textit{variational inference} (VI), which allows for scalable posterior approximation in a setting with tens of thousands of latent variables \parencite{blei2017variational,kucukelbir2017automatic}.

Against this background, the paper asks whether a dynamic factor model with inertia can jointly achieve predictive accuracy, efficient representation of heterogeneity through cross-category pooling and scalable, reliable  Bayesian inference in high-dimensional retail environments.

A simulation study calibrated to realistic retail environments evaluates predictive performance, scalability and economic implications. The dynamic factor specification outperforms both static factor models and mixed logit benchmarks in out-of-sample prediction, with performance gains that increase as the number of categories expands. Incorporating inertia also meaningfully alters implied own-price elasticities, highlighting that dynamics are consequential not only for predictive accuracy, but also for substantive economic conclusions.

The remainder of the paper proceeds as follows. Section \ref{s: model} outlines the consumer choice environment and introduces the dynamic product-level factor structure designed to mitigate the curse of dimensionality. Section \ref{s: est} details the Bayesian estimation strategy and the VI procedure used to approximate the posterior distribution. Section \ref{s: sim_study} presents simulation evidence on predictive performance, scalability and the resulting elasticity estimates. Section \ref{s: conclusion} concludes and discusses avenues for future research.

\section{Model}\label{s: model}

\subsection{Consumer Choice Framework}\label{s: cons_choice}
Consumer choice is modeled at the product level in a multi-category setting with repeated purchases. We observe consumers $i\in\{1,\dots,I\}$ making choices on shopping trips $t\in\{1,\dots,T\}$. On each trip, consumer $i$ potentially purchases one product from a subset of categories $c\in\{1,\dots,C\}$. Conditional on purchasing in category $c$ at trip $t$, consumer $i$ chooses exactly one product from the set $\mathcal{J}_c \equiv\{1,\dots,J_c\}$.

We adopt a random-utility framework \parencite{mcfadden1974conditional,train2009discrete}. The utility from choosing product $j_c\in\mathcal{J}_c$ in category $c$ at trip $t$ is
\begin{align}
    U_{ij_ct}=u_{ij_ct} + \epsilon_{ij_ct},
\end{align}
where $u_{ij_ct}$ is a deterministic component and $\epsilon_{ij_ct}$ is an idiosyncratic shock. We assume $\epsilon_{ij_ct}$ is i.i.d. Gumbel distributed, so that conditional choice probabilities take the multinomial logit form. Each consumer chooses the product that provides the highest utility among all alternatives in a category
\begin{align} \label{eq: y_ijt}
    y_{ij_ct} = 1 \quad \text{if} \quad j_c = \underset{k\, \in \, \mathcal{J}_c}{\text{argmax}} \, U_{ikt}.
\end{align}

The deterministic component of utility may depend on baseline preferences, a vector of observed time-varying product characteristics and on the history of past choices. Let $\mathbf{x}_{j_ct}$ denote observed covariates such as price, promotions or other product attributes and let $\mathcal{H}_{i,t-\ell}\equiv \{X_{i,t-1}, X_{i,t-2}, \dots, X_{i,t-T} \}$ summarize the history of consumer $i$'s past choices and states. A general dynamic specification can be written as 
\begin{align} \label{eq: general_dyn_u}
u_{ij_ct}
= \alpha_{ij_c} + \boldsymbol{\zeta}_{ij_c}' \mathbf{x}_{j_ct} + f\big(\mathcal{H}_{i,t-\ell}\big),
\end{align}
where $\boldsymbol{\zeta}_{ij_c}$ captures heterogeneous responses to observed covariates and $f(\cdot)$ is an arbitrary function mapping past history into current utility.

Our baseline specification imposes two simplifying restrictions that enable large scale estimation. We restrict contemporaneous covariate effects to be linear in price and model dynamics using one-period within-category state dependence.\footnote{For the purposes of the simulation, we rely on this simplified specification, noting that the framework can be extended to incorporate additional product characteristics and consumer demographics, as in \textcite{Donnelly2021counterfactual}, and to allow for richer history dependence $f\big(\mathcal{H}_{i,t-\ell}\big)$.}
Let $\delta_{ij_c, t-1}$ be an indicator that equals one if consumer $i$ chose product $j_c$ on the most recent previous purchase occasion in category $c$ and zero otherwise. The baseline specification is
\begin{align} \label{eq: det_u}
    u_{ij_ct} = \alpha_{ij_c} - \eta_{ij_c}\,\mathrm{price}_{j_ct} + \xi_{ij_c}\, \delta_{ij_c,t-1},
\end{align}
which corresponds to the special case of equation \eqref{eq: general_dyn_u} in which $\boldsymbol{\zeta}_{ij_c}' \mathbf{x}_{j_ct} = - \eta_{ij_c}\,\mathrm{price}_{j_ct}$ and $f\big(\mathcal{H}_{i,t-\ell}\big) = \xi_{ij_c}\, \delta_{ij_c,t-1}$. 
This specification captures (i) heterogeneous baseline preferences in $\alpha_{ij_c}$, (ii) heterogeneous price sensitivity in $\eta_{ij_c}$ and (iii) heterogeneous inertia in $\xi_{ij_c}$. A fully flexible product-level model would treat these as unrestricted consumer-product coefficients. However, this requires the estimation of $3\sum_c I\cdot J_c$ parameters, which is infeasible in large-scale retail applications.

A natural way to constrain the dimensionality of \eqref{eq: det_u} is to impose structure on the heterogeneous coefficients. We therefore adopt a mixed logit model \parencite{mcfadden2000mixed} as a benchmark with the deterministic utility component defined as
\begin{align}\label{eq: mix_log}
    u_{ij_ct} = \alpha_{j_c} - \eta_{i} \, \mathrm{price}_{j_ct} + \xi_{i} \, \delta_{ij_c,t-1}.
\end{align}

Baseline product attractiveness is captured by product fixed effects $\alpha_{j_c}$, which absorb average differences in popularity within each category and are homogeneous across consumers. For tractability, consumer heterogeneity is restricted to price sensitivity and inertia through random coefficients $(\eta_i, \xi_i)$. This benchmark substantially reduces the number of heterogeneous parameters relative to the fully flexible product-level specification while preserving heterogeneity in price responsiveness and inertia.

We focus on product-level demand conditional on category choice and abstract from modeling higher-level choice margins. The framework can be extended to incorporate category choice through a nested or hierarchical structure, as in \textcite{Donnelly2021counterfactual}, allowing substitution patterns to operate both within and across categories. An additional upper layer could model store or supermarket choice, yielding a mulit-level demand system that jointly captures retailer, category and product decisions.

\subsection{Dynamic Factor Structure}\label{s: dyn_fac}
Another approach to mitigate the curse of dimensionality is to put a factor structure on the heterogeneity so that information can be shared across products and categories while preserving rich substitution patterns \parencite{Ruiz2020shopper, Donnelly2021counterfactual}. The core idea is to represent the full matrix of consumer-product coefficients by the inner product of a shared $K$-dimensional latent consumer vector and product specific loading vectors \parencite{salakhutdinov2008probabilistic,koren2009matrix}.

We posit that each consumer $i$ has a latent vector $\boldsymbol{\theta}_i \in \mathbb{R}^K$ that is shared across all categories. It represents all relevant unobserved heterogeneity. Each product $j_c$ is associated with three loading vectors: $\boldsymbol{\gamma}_{j_c} \in \mathbb{R}^K$ for baseline utility, $\boldsymbol{\lambda}_{j_c} \in \mathbb{R}^K$ for price sensitivity and $\boldsymbol{\rho}_{j_c} \in \mathbb{R}^K$ for inertia.
Since the same dimensions of $\boldsymbol{\theta}_i$ are allowed to affect all these aspects of choice, this gives our model a lot of flexibility if the dimension $K$ is large enough. For example, certain households might have a relatively high price sensitivity and a low inertia at the same time. 
Substituting the latent factor structure in equation \eqref{eq: det_u} yields the dynamic factor model used throughout the paper:
\begin{align}
     u_{ij_ct} = \boldsymbol{\theta}_i' \boldsymbol{\gamma}_{j_c} - \boldsymbol{\theta}_i'\boldsymbol{\lambda}_{j_c}\,\mathrm{price}_{j_ct} + \boldsymbol{\theta}_i'\boldsymbol{\rho}_{j_c}\,\delta_{ij_c,t-1}.
\end{align}
This factorization reduces the number of parameters to be estimated from $O\big(I\cdot \sum_cJ_c \big)$ to $O\big(IK + K \cdot \sum_c J_c\big)$ per coefficient block and it induces correlated heterogeneity across products and categories through the shared $\boldsymbol{\theta}_i$.

Conditional on purchasing any product $j$ in category $c$ at trip $t$, the i.i.d. Gumbel assumption about $\epsilon_{ij_ct}$ implies multinomial logit choice probabilities \parencite{mcfadden1974conditional}:
\begin{align} \label{eq: choice_prob}
    p\Big(y_{ij_ct}=1 \Big| \sum_{j_c = 1}^{J_c} y_{ij_ct} = 1\Big) =  \frac{exp(u_{ij_ct})}{\sum_{m_c=1}^{J_c} exp(u_{im_ct})},
\end{align}
where $y_{ij_ct} \in \{0, 1\}$ denotes the binary outcome variable indicating whether consumer $i$ chooses product $j_c$ in category $c$ on trip $t$. 

Note that for the initial observation for each consumer, we cannot calculate the dynamic choice probabilities since the lagged dependent variables $\delta_{ij_c0}$ are unobserved. This problem and solutions corresponds to the initial conditions problem of general discrete choice models \parencite{heckman1981}. In the simulation study in Section \ref{s: sim_study}, we abstract from this by exogenously generating initial choices.

The conditional log-likelihood is 
\begin{align}\label{eq: log_lik}
    log(\mathcal{L}) = \sum_{i=1}^{I}\sum_{c=1}^{C}\sum_{t=1}^{T} \sum_{j_c=1}^{J_c} y_{ij_ct}\left[ u_{ij_ct} - \log\!\left(\sum_{m_c=1}^{J_c} \exp(u_{im_ct})\right)
\right].
\end{align}

Because latent factor models are invariant to several transformations, identification requires normalizations \parencite{anderson1956statistical}. In the present model, four such invariances arise. 
Scale invariance arises because utilities depend on inner products between individual factors and product loadings. For any nonzero scalar $a$, multiplying all individual factors by $a$ and dividing all loading vectors by $a$ leaves the inner products $\boldsymbol{\theta}_i' \boldsymbol{\gamma}_{j_c}, \boldsymbol{\theta}_i'\boldsymbol{\lambda}_{j_c}$ and $\boldsymbol{\theta}_i'\boldsymbol{\rho}_{j_c}$ unchanged. We fix the scale by imposing standard normal priors on individual factors, $\theta_{ik} \sim\mathcal{N}(0,1)$ in our Bayesian estimation approach. This normalization fixes the variance of each latent dimension, thereby eliminating the multiplicative invariance since any rescaling would violate the prior variance restriction.

Sign invariance is a special case of scale invariance. Utilities remain unchanged if the sign of a factor and its associated loadings is flipped. Formally, for any factor index $k$, the transformation $(\theta_{ik}, \gamma_{kj_c},\lambda_{kj_c},\rho_{kj_c})\mapsto(-\theta_{ik}, -\gamma_{kj_c},-\lambda_{kj_c},-\rho_{kj_c})$ leaves utility unchanged. To eliminate this invariance, we impose an anchor restriction by constraining one loading per factor (e.g., the first product in each category) to be strictly positive.

Because the model does not include an explicit outside option, it suffers from location invariance. Conditional choice probabilities depend only on within-category utility differences. Adding a category-specific constant to all utilities therefore does not affect observed choices. We remove this arbitrary utility level by mean-centering utilities within each category.

Rotational invariance arises because only the latent $K$-dimensional subspace is identified, not its particular basis representation. For any orthogonal matrix $Q$, the transformation $$(\theta_{ik}, \gamma_{kj_c},\lambda_{kj_c},\rho_{kj_c})\mapsto(Q\theta_{ik}, Q\gamma_{kj_c},Q\lambda_{kj_c},Q\rho_{kj_c})$$ preserves all inner products and thus utilities.
We do not impose additional restrictions to resolve rotational invariance, as our analysis focuses on prediction and elasticity estimation rather than structural interpretation of individual factor dimensions \parencite{lopes2004factor}. Consequently, loadings are identified only up to orthogonal rotation.

\section{Estimation}\label{s: est}

Following \textcite{Donnelly2021counterfactual}, we estimate the dynamic factor model from section \ref{s: dyn_fac} in a Bayesian framework. Let $X$ denote the observed data, consisting of past choices and prices, and let
\begin{align}
    \boldsymbol{\beta} \equiv\{\boldsymbol{\theta}_i\}_{i=1}^I \cup \{\boldsymbol{\gamma}_{j_c}, \boldsymbol{\lambda}_{j_c}, \boldsymbol{\rho}_{j_c}\}_{c=1, j\in \mathcal{J}_c}^C
\end{align}
collect all latent parameters. 
The goal is to estimate the posterior distribution of the parameter vector $\boldsymbol{\beta}$, which by Bayes' rule satisfies
\begin{align}
    p(\boldsymbol{\beta}\, |\, X) \propto p(\boldsymbol{\beta})\, p(X\,|\,\boldsymbol{\beta}).
\end{align}
Here, $p(X\,|\,\boldsymbol{\beta})$ is the likelihood function defined in equation \eqref{eq: log_lik} and $p(\boldsymbol{\beta})$ denotes the prior distribution.

The posterior $p(\boldsymbol{\beta}\, |\, X)$ does not admit a closed-form representation. The log-likelihood function contains nonlinear log-sum-exp terms that couple a large number of latent variables across individuals and products, rendering posterior moments analytically intractable. Moreover, exact \textit{Markov Chain Monte Carlo} (MCMC) methods are computationally infeasible at the scale considered here \parencite{Betancourt2018conceptual}. We therefore rely on VI to obtain a scalable approximation to the posterior distribution \parencite{blei2017variational}.

VI replaces posterior simulation by optimization. Specifically, we select a tractable family of distributions $q(\boldsymbol{\beta}\,;\, \nu)$, indexed by variational parameters $\nu$ and select  
\begin{align}
    \nu ^* = \text{arg}\, \underset{\nu}{\text{min}} \,KL\big(q(\boldsymbol{\beta} \, ;\,\nu) \,\|\, \, p(\boldsymbol{\beta}\, | \,X)\big)
\end{align}
to minimize the \textit{Kullback-Leibler} (KL) divergence. This is equivalent to maximizing the \textit{evidence lower bound} (ELBO),
\begin{align}
    \mathcal{L}(\nu) = \mathbb{E}_{q(\boldsymbol{\beta};\nu)}\big[\log p(X\, |\,\beta) - \log q(\boldsymbol{\beta}\,;\nu)\big],
\end{align}
where $\mathbb{E}_{q(\boldsymbol{\beta};\nu)}$ denotes the expectation with respect to $q(\boldsymbol{\beta};\nu)$. 
We adopt a mean-field Gaussian variational family that factorizes across individuals and products
\begin{align}
    q(\boldsymbol{\beta};\nu) 
    = \prod_{i=1}^I q(\boldsymbol{\theta}_i;\nu_{\theta_i}) 
      \prod_{c=1}^C \prod_{j_c=1}^{J_c} 
      q(\boldsymbol{\gamma}_{j_c};\nu_{\gamma_{j_c}}) \,
      q(\boldsymbol{\lambda}_{j_c};\nu_{\lambda_{j_c}}) \,
      q(\boldsymbol{\rho}_{j_c};\nu_{\rho_{j_c}}),
\end{align}
where each block parameter is modeled as a multivariate normal distribution with diagonal covariance matrix.
Concretely, for each $K$-dimensional parameter vector $\boldsymbol{z}\in\{\boldsymbol{\theta}_i,\boldsymbol{\gamma}_{j_c},\boldsymbol{\lambda}_{j_c},\boldsymbol{\rho}_{j_c}\}$ the variational distribution is
\begin{align}
  q(\boldsymbol{z})=\mathcal N\bigl(\mu_{\boldsymbol{z}},\operatorname{diag}(\sigma_{\boldsymbol{z}}^2)\bigr),
\end{align}
so that the variational parameters $\nu$ collect all means and log standard deviations.

Optimization is performed using \textit{automatic differentiation variational inference} (ADVI) \parencite{kucukelbir2017automatic}, which use Monte Carlo integration to approximate the ELBO. Using the reparameterization trick \parencite{kingma2014autoencoding}, draws are written as $z=\mu_z+\sigma_z\odot\epsilon$ with $\epsilon\sim\mathcal N(0,I)$, which yields low-variance stochastic gradient estimators of the ELBO. Gradients are computed by automatic differentiation and updated using stochastic gradient ascent \parencite{hoffman2013stochastic,ranganath2014blackbox}. Convergence is monitored via the mean change in ELBO.

The mean-field factorization imposes posterior independence across parameter vectors and restricts covariance matrices to be diagonal. While this assumption substantially improves scalability in high-dimensional settings, it typically understates posterior uncertainty by ignoring cross-parameter correlations. In the present application, the primary objectives of interest are predictive choice probabilities and implied elasticities, which depend mainly on posterior means rather than higher-order dependence structures. The mean-filed approximation therefore offers a computationally efficient trade-off between tractability and accuracy in a model with tens of thousands of latent parameters.

A common concern in demand estimation is price endogeneity. If prices respond to unobserved demand shocks, they may be correlated with the unobserved component of utility, biasing estimates of price sensitivity. In empirical applications, this issue can be addressed through \textit{instrumental variables} (IV). In a Bayesian framework, IV estimation can be implemented by specifying a joint model of demand and pricing, allowing for correlation between their structural disturbances \parencite{kleibergen2003bayesian}. This approach integrates over the joint posterior of demand and pricing parameters and propagates first-stage uncertainty directly into the demand estimates.

In the following simulation study, price endogeneity is not a concern by construction. Prices are generated exogenously from a stochastic process that is independent of the utility shocks. Consequently, the identifying variation in prices is orthogonal to unobserved demand components and price coefficients are consistently estimated without the need for IVs. In empirical applications, however, additional identifying assumptions or instruments would be required. 

\section{Simulation Study}\label{s: sim_study}
\subsection{Data Generating Process}\label{s: dgp}
We simulate repeated purchase data intended designed to resemble retail scanner settings with many products, multiple categories and sparse individual purchase histories.
Let $i \in \{1, \dots, I\}$ index individuals, $c\in\{1,\dots, C\}$ categories, $j_c\in \{1, \dots, J_c\}$ products within category $c$ and $t\in\{1, \dots, T\}$ shopping trips per individual.

Preferences follow a K-dimensional latent factor structure. For each individual and product, we draw
\begin{align}
\boldsymbol{\theta}_i \sim \mathcal U(0,1)^{K},\quad
    \boldsymbol{\gamma}_{j_c} \sim \mathcal U(0.3,0.4)^{ K},\quad
    \boldsymbol{\lambda}_{j_c} \sim \mathcal U(0.3,0.4)^{K}, \quad
    \boldsymbol{\rho}_{j_c} \sim \mathcal{U}(0,2)^K.
\end{align}

These draws generate heterogeneous baseline utilities $\boldsymbol{\theta}_i'\boldsymbol{\gamma}_{j_c}$, heterogeneous price sensitivities $\boldsymbol{\theta}_i'\boldsymbol{\lambda}_{j_c}$ and heterogeneous inertia strengths $\boldsymbol{\theta}_i'\boldsymbol{\rho}_{j_c}$.
The number of factors $K$ controls the dimensionality and correlation structure of heterogeneity.

Uniform distributions are used for the latent factors and loadings to generate flexible but controlled heterogeneity without imposing additional structure on higher-order moments. In particular, uniform draws avoid mechanically inducing normality or specific tail behavior in utilities, allowing the simulation to isolate the role of dimensionality, correlation induced by the factor structure and inertia strength. Because the focus of the simulation is comparative performance across model specifications rather than recovering a particular parametric distribution, uniform draws provide a transparent and design-neutral benchmark.

Prices follow a two-level process with category-specific scales and trip-specific shocks.
For each category $c$, we draw a baseline level $m_c$ and a dispersion parameter $r_c$
\begin{align}
    m_c\sim\mathcal U(1,10), \quad r_c\sim\mathcal U(0,1),
\end{align}
and set baseline product prices $p_{j_c}\sim\mathcal U(m_c,m_c+r_c)$.
At each trip $t$, prices are affected by idiosyncratic shocks $\varepsilon_{j_ct}\sim\mathcal N(0,0.2)$ and truncated below at $0.1$, yielding 
\begin{align}
    p_{j_ct}=\max\bigl(0.1,\ p_{j_c}+\varepsilon_{j_ct}\bigr).
\end{align}

To generate multi-category baskets and realistic sparsity, category incidence is modeled separately from product choice.
At each trip $t$, individual $i$ purchases in $C_{it}\sim\mathcal U(\lfloor\frac{C}{2}\rfloor,C)$ categories and a subset $\mathcal C_{it}\subset\{1,\dots,C\}$ of that size is selected.
Conditional on $c \in \mathcal{C}_{it}$, individual $i$ chooses the product that maximizes his utility
\begin{align}
     U_{ij_ct} = \boldsymbol{\theta}_i' \boldsymbol{\gamma}_{j_c} - \boldsymbol{\theta}_i'\boldsymbol{\lambda}_{j_c}\,p_{j_ct} + \boldsymbol{\theta}_i'\boldsymbol{\rho}_{j_c}\,\delta_{ij_c,t-1} + \epsilon_{ij_ct},
\end{align}
with $\epsilon_{ij_ct} \sim \text{Gumbel}(0,1)$ this yields standard multinomial logit choice probabilities, as in equation \eqref{eq: choice_prob}.
We draw $y_{ij_ct}$ according to equation \eqref{eq: y_ijt} and update $\delta_{ij_c, t-1}$. In the simulation, inertia is exogenous by construction.

\begin{figure}[H]
    \centering
    \includegraphics[width=\linewidth]{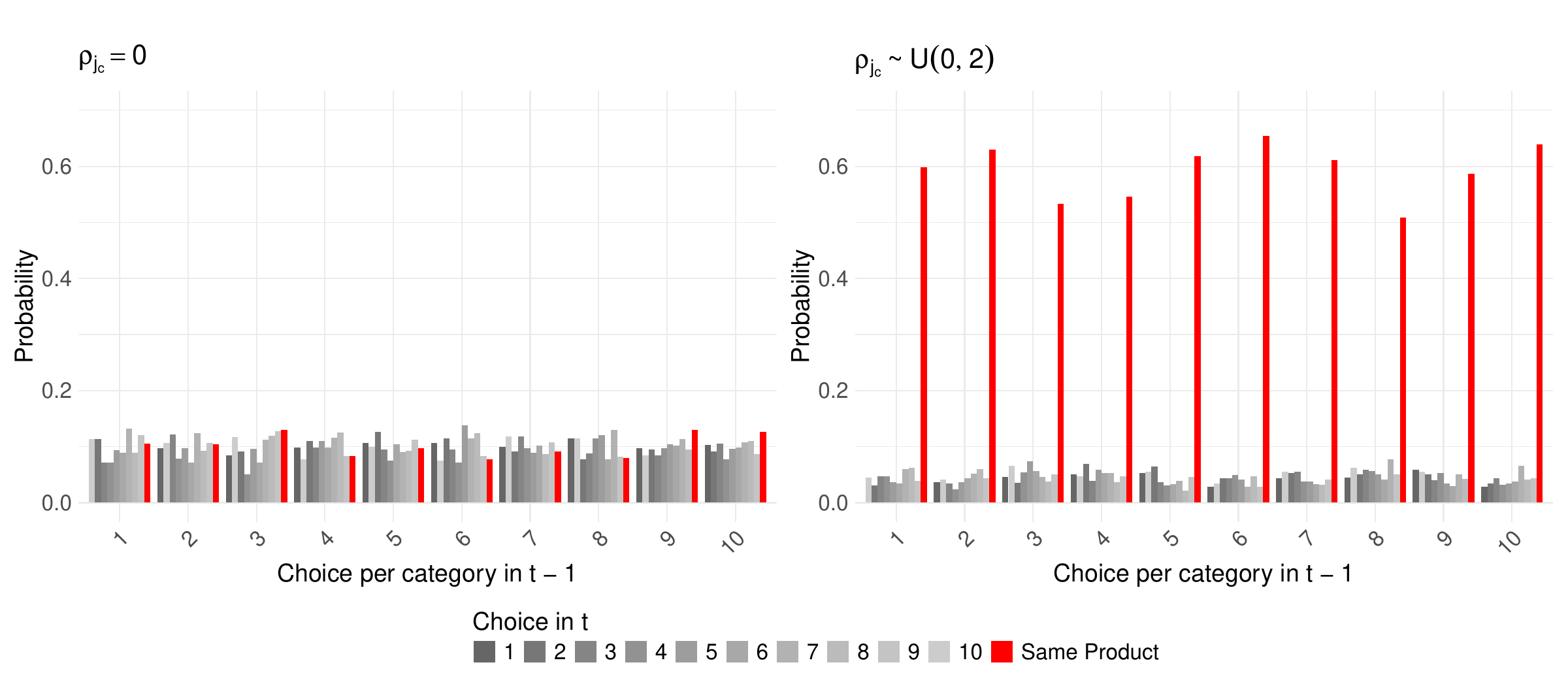}
    \caption{Probability of choosing the same product in $t$ as in $t-1$ or switching to another product in a setting with $C=10$ categories and $J_c =10$ products per category.}
    \label{fig: dgp_inertia}
\end{figure}

Figure \ref{fig: dgp_inertia} illustrates the effect of the inertia in a setting with $C=10$ categories and $J_c=10$ product per category. When $\boldsymbol{\rho}_{j_c}=0$, choice probabilities are approximately independent of the previously purchased product. When $\boldsymbol{\rho}_{j_c}\sim \mathcal{U}(0,2)$, the probability of repurchasing the same product increases drastically, generating strong persistence in observed choices while preserving the possibility to switch to another product.

\subsection{Results}\label{s: results}
This section evaluates the predictive and economic performance of the dynamic product-level factor model. While the design favors a factor structure with inertia, the results provide insight into how the model behaves as dimensionality, sparsity and sample size vary.

We implement the model in probabilistic programming language \texttt{Stan} \parencite{carpenter2017stan} with standard normal priors $\mathcal N(0,1)$ on all latent variables (as in \textcite{Donnelly2021counterfactual}) and approximate the posterior using VI as described in Section \ref{s: est}.\footnote{A robustness analysis using a less informative prior specification is reported in Appendix \ref{a: prior_check}.}
All computations are executed on the Hilbert Cluster at Heinrich Heine University D\"usseldorf.
For each experimental setting we generate $R=25$ replication datasets and split each dataset into $80\%$ training and $20\%$ test data.
We retain $S=4000$ posterior samples and compute predicted choice probabilities $\hat p_{s,n,j_c}$ for each posterior draw $s \in S$ and observation in the test set $n \in N$.
Because the \textit{data generating process} (DGP) is known, predictive performance can be evaluated against the true choice probabilities $p^{\mathrm{true}}_{n,j_c}$.
We summarize prediction accuracy by the \textit{root mean squared error} (RMSE) for each draw $s$
\begin{align}
  \mathrm{RMSE}_s
  =\sqrt{\frac{1}{N\cdot J_c}\sum_{n=1}^{N}\sum_{j\in\mathcal J_c}\bigl(\hat p_{s,n,j_c}-p^{\mathrm{true}}_{n,j_c}\bigr)^2}
\end{align}
and report it as the mean across all posterior draws and replications.
As a benchmark, we estimate the mixed logit specification from equation \eqref{eq: mix_log} using the \texttt{mlogit} package in \texttt{R} \parencite{croissant2020mlogit} and a static factor model without the inertia term $\boldsymbol{\theta}_i'\boldsymbol{\rho}_{j_c}\,\delta_{ij_c,t-1}$ using the same estimation approach as for the dynamic factor model.

\begin{figure}[H]
  \centering
  \includegraphics[width=\linewidth]{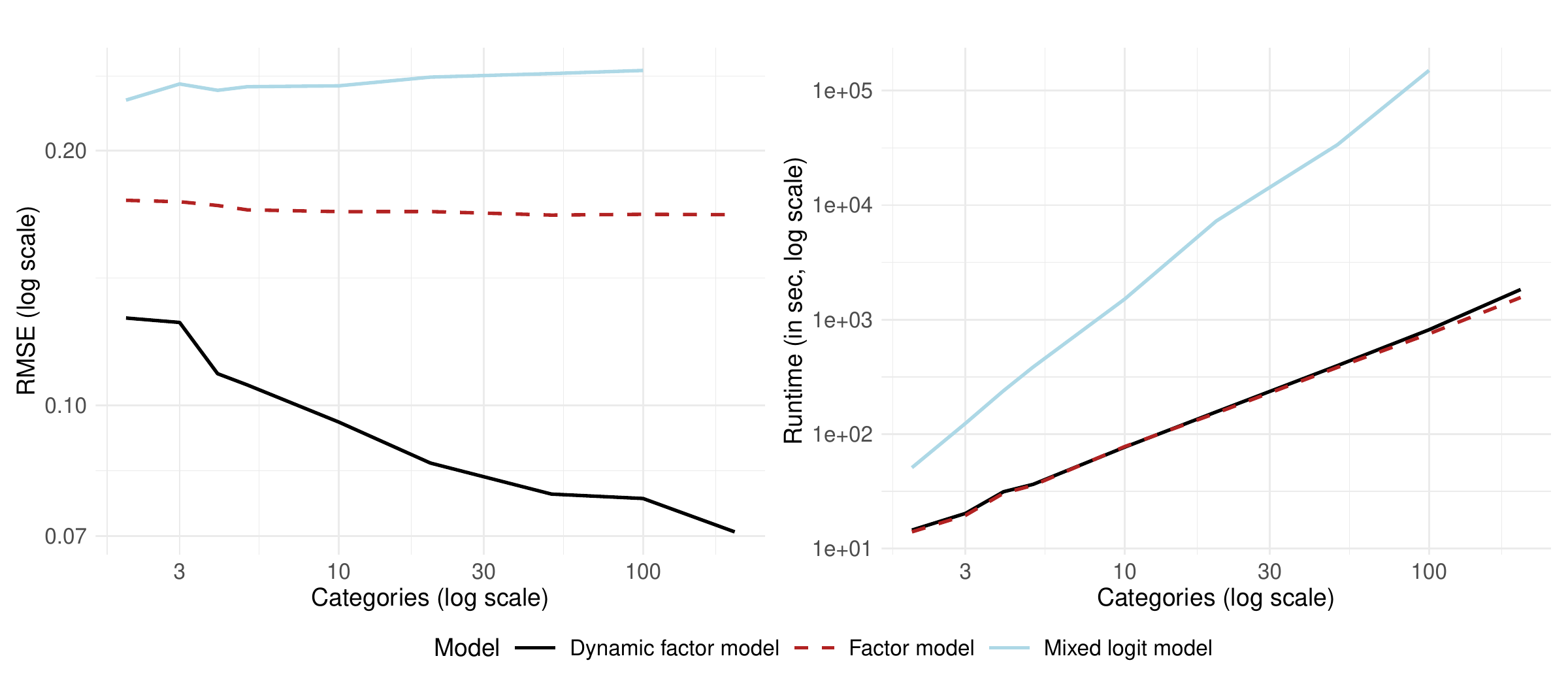}
  \caption{Simulation results for up to $C=200$ categories with $R=25$ datasets per setting and $I=40$, $J=10$, $T=20$ and $K=5$.}
  \label{fig:sim_results_categories}
\end{figure}

Figure \ref{fig:sim_results_categories} reports test-set RMSE (left panel) and runtime (right panel) as the number of categories increases up to $C=200$ (holding $I=40$, $J=10$, $T=20$ and $K=5$ fixed).
In the left panel, three patterns emerge. First, the RMSE of the dynamic factor model declines as the number of categories increases. Although the dimensionality of the choice problem expands, additional categories provide informative cross-category variation that sharpens inference on the shared latent factors an inertia parameters. Because the factor structure links preferences across categories, observed behavior in one market helps identify tastes that are relevant elsewhere. In this design, the informational gains from cross-category pooling outweigh the increase in dimensionality.

Second, the static factor model, which omits inertia but retains the shared latent structure, exhibits relatively stable RMSE as the number of categories grows. The factorization continues to pool information across categories, preventing decline in predictive accuracy. However, because the DGP includes state dependence, the static model remains misspecified. Additional categories do not considerably reduce this structural misspecification and predictive performance remains approximately constant.

Third, the mixed logit model displays increasing RMSE as the number of categories expands.\footnote{We cannot report RMSE for the mixed logit model with $C=200$ categories due to runtime restrictions on the cluster.} Unlike the factor specifications, the mixed logit model does not impose a shared low-dimensional structure across categories and is estimated separately by category. As the number of categories increases, each category contributes limited within-category information and the absence of cross-category pooling lead to progressively weaker identification and rising prediction error.

Taken together, the results indicate that the shared factor structure becomes particularly valuable in high-dimensional environments. The dynamic specification not only remains stable but benefits from additional cross-category information.

The right panel of figure \ref{fig:sim_results_categories} reports runtime on a log scale. The dynamic and static factor models exhibit roughly proportional growth in runtime  as the number of categories increases, consistent with the need to evaluate multinomial logit choice probabilities over a large dataset.
By contrast, the mixed logit model exhibits a sharply accelerating runtime profile. The increase is close to exponential in the number of categories, reflecting the simulation-based integration required for random coefficients and the absence of a shared low-dimensional representation. As dimensionality grows, computational cost rises disproportionately relative to the factor-based specifications.

These results highlight a dual advantage of the dynamic factor approach in this simulation design: improved predictive performance and favorable computational scaling as the number of categories increases. At the same time, the interpretation should remain thoughtful. The DGP follows a dynamic factor structure with inertia, so the dynamic model is correctly specified, while the static and mixed logit benchmarks are misspecified. The observed divergence therefore reflects performance under specification alignment. The extent to which similar gains arise under alternative data-generating environments remains an empirical question.

Table \ref{tab:model_performance} complements Figure~\ref{fig:sim_results_categories} by reporting test-set RMSE of the dynamic factor model, accuracy and runtime across variations in the cross-sectional dimension $I$, purchase-history length $T$, number of products per category $J$ and number of categories $C$. In contrast to the RMSE, the accuracy is measured as the ratio of correctly predicted product choices therefore evaluating discrete classification performance instead of probabilistic calibration.

Increasing the number of individuals from $I=40$ to $I=1000$ substantially improves predictive performance. For example, at $J=10$, $T=5$ and $C=10$, RMSE falls from $0.1629$ to $0.0795$ and accuracy rises from $0.2692$ to $0.4372$. Similar improvements are observed across all settings. This improvement reflects that a richer cross section sharpens inference on the shared latent structure.

Longer purchase histories improve both RMSE and accuracy. This pattern is consistent with the interpretation of inertia as state dependence that becomes more informative when more within-category transitions are observed and it aligns with classic identification arguments that longer panels help distinguish true state dependence from persistent heterogeneity \parencite{heckman1981}.

Increasing the number of products per category affects the two metrics differently. When $J$ rises from $10$ to $100$ (holding $I=40$, $T=20$ and $C=10$ fixed), RMSE declines from $0.0874$ to $0.0232$, while accuracy drops sharply from $0.5295$ to $0.0378$. The decline in accuracy reflects the inherent difficulty of predicting the exact chosen alternative in a large choice set. Even well-calibrated models will exhibit low classification rates when many alternatives are available. RMSE, by contrast, benefits mechanically from smaller per-product choice probabilities in larger choice sets.
Finally, runtime increases with the number of observations and with the size of the choice problem (in particular, with $I$, $C$ and $T$), which is the expected scaling given the need to evaluate choice probabilities repeatedly during inference.

\begin{table}[H]
  \centering
  \small
  \setlength{\tabcolsep}{5pt}
  \begin{tabular}{cccc|rr|cc}
    \hline
    \multicolumn{4}{c|}{\textit{Setting}} &  & Runtime &  &  \\
    $I$ & $J$ & $T$ & $C$ & Observations & (sec) & RMSE & Accuracy \\
    \hline
    40 & 10 & 5  & 10  & 1193   & 33    & 0.1629 & 0.2692 \\
    1000 & 10 & 5 & 10 & 29924  & 782   & 0.0795 & 0.4372 \\
    40 & 10 & 5  & 100 & 11995  & 399   & 0.1519 & 0.2252 \\
    1000 & 10 & 5 & 100 & 299851 & 8407 & 0.0501 & 0.4565 \\
    40 & 10 & 20 & 10  & 4791  & 125   & 0.0847 & 0.5295 \\
    1000 & 10 & 20 & 10 & 119744 & 3108 & 0.0516 & \textbf{0.5792} \\
    40 & 100 & 20 & 10  & 4800  & 1165  & 0.0232 & 0.0378 \\
    1000 & 100 & 20 & 10 & 119652 & 26514 & \textbf{0.0132}& 0.1325 \\
    \hline
  \end{tabular}
  \caption{Dynamic factor model: results averaged over $R=25$ replications per setting; RMSE and accuracy are reported on the test data (rounded).}
  \label{tab:model_performance}
\end{table}

We evaluate the economic implications of the estimated demand system by computing own-price elasticities from posterior draws of the dynamic factor model. Own-price elasticities are obtained by partially differentiating the multinomial logit choice probability \eqref{eq: choice_prob} with respect to the product's own price and scaling by the ratio of its price to choice probability. For consumer $i$, product $j_c$, category $c$, trip $t$ and posterior draw $s$, the own-price elasticity is given by
\begin{align}
\mathcal E_{ij_ct}^{own,(s)} = -\left(\theta_i^{(s)\prime}\lambda_{j_c}^{(s)}\right)
\ \text{price}_{j_ct}
\left(1-P_{ij_ct}^{(s)}\right),
\end{align}
where \(P_{ij_ct}^{(s)}\) denotes the predicted choice probability.
We then aggregate these draw-specific elasticities across observations within each product-category pair. Posterior means summarize average elasticities, while credible intervals are constructed from the empirical posterior quantiles across draws.

\begin{figure}[H]
  \centering
  \includegraphics[width=\linewidth]{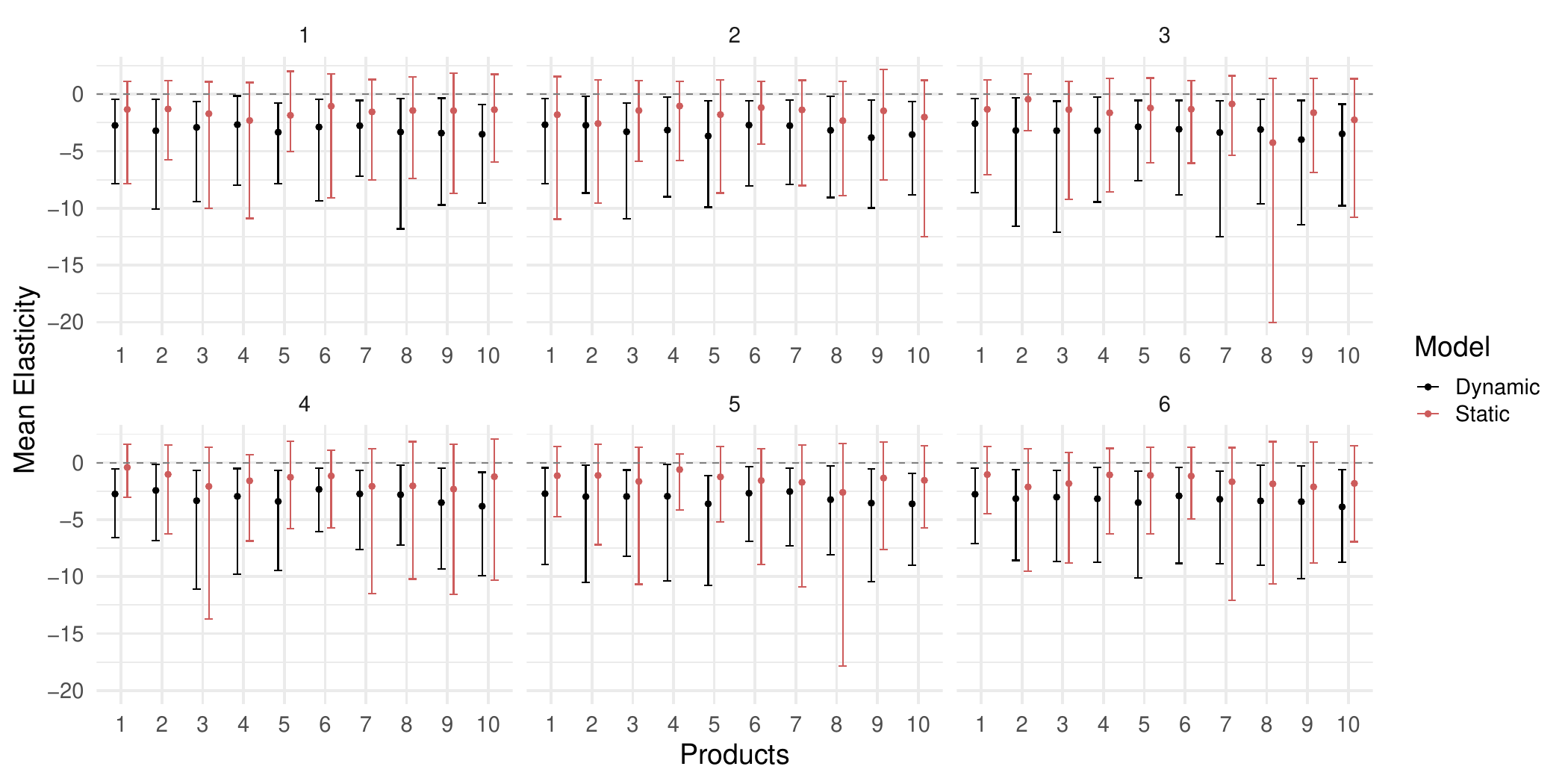}
  \caption{Mean own-price elasticities with $95\%$ credibility intervals by category for $C=6$ categories with $J_c=10$ products per category.}
  \label{fig:elasticities_by_category}
\end{figure}

Figure \ref{fig:elasticities_by_category} reports mean own-price elasticities by category together with $95\%$ credibility intervals for both the dynamic and static specifications over $6$ categories with $J_c=10$ products in each category. Several observations emerge.

First, the dynamic specification yields uniformly negative mean own-price elasticities, with magnitudes that vary across products and categories. This heterogeneity reflects both realized price dispersion and differences in estimated price sensitivity across latent dimensions.

Second, the static specification produces noticeably wider credibility intervals and importantly, the reported 95\% credibility intervals for the static elasticities include zero across all products. In a classical frequentist interpretation, this would suggest that own-price effects are not statistically distinguishable from zero at conventional levels. In the present Bayesian framework, however, the intervals reflect posterior uncertainty rather than sampling-based hypothesis tests. The fact that zero lies within the posterior interval indicates that the data provide limited information about price responsiveness under the static specification, particularly once inertia is omitted.

This pattern is economically informative. When inertia is excluded, persistent purchase behavior may be attributed to unobserved heterogeneity rather than state dependence, weakening the estimated relationship between price changes and switching behavior. As a result, the static model generates elasticities that are both more weakly identified and more uncertain. The dynamic specification, by explicitly modeling state dependence, appears to sharpen inference on price sensitivity. Again, these differences should be interpreted cautiously. Because the DGP includes inertia by construction, the broader confidence intervals and near-zero elasticities in the static case partly reflect structural misspecification rather than purely statistical limitations.

\section{Conclusion}\label{s: conclusion}
Large-scale retail environments combine high-dimensional product assortments with sparse purchase histories at the individual level.
This paper proposes a dynamic product-level factor model that addresses these features by combining a shared latent factor structure for heterogeneity with within-category state dependence.
In simulation designs calibrated to realistic scanner-data settings, the model delivers improved predictive accuracy relative to static factor specifications and mixed logit benchmarks, with gains that become more pronounced as the number of categories increases.
The economic implications are substantive: accounting for inertia changes the distribution of inferred price sensitivities and therefore affects implied demand elasticities.

Several directions for future research appear promising.
First, richer inertia mechanisms could be incorporated to better capture consumer dynamics.
Examples include category-level switching costs, multiple-lag dependence and state variables that summarize longer purchase histories, as well as formulations that allow inertia to vary with observed marketing variables such as promotions.
Second, applying the framework to real-world scanner data would require addressing additional empirical challenges, including price endogeneity, unobserved choice sets and the modeling of category incidence (the decision to purchase in a category) alongside within-category brand choice.
Third, the factor structure could be extended to allow time-varying preferences or hierarchical priors that link products within and across categories.
Taken together, these extensions would further strengthen the case for dynamic factor models as a scalable approach to demand estimation in modern retail markets.

\newpage
\pagestyle{plain}

\renewcommand*{\bibfont}{\small}

\printbibliography 

@article{Athey2018ttfm,
	author = {Athey, S. and Blei, D. and Donnelly, R. and Ruiz, F. and Schmidt, T.},
	title = {Estimating Heterogeneous Consumer Preferences for Restaurants and Travel Time Using Mobile Location Data},
	journal = {AEA Papers and Proceedings},
	volume = {108},
	year = {2018},
	pages = {64-67}
}

@article{Donnelly2021counterfactual,
	title={Counterfactual inference for consumer choice across many product categories},
	author={Donnelly, R. and Ruiz, F. J. R. and Blei, D. and Athey, S.},
	journal={Quantitative Marketing and Economics},
	pages={369--407},
	volume={19},
	year={2021}
}

@article{kucukelbir2017automatic,
	title={Automatic differentiation variational inference},
	author={Kucukelbir, A. and Tran, D. and Ranganath, R. and Gelman, A. and Blei, D. M.},
	journal={Journal of Machine Learning Research},
	volume={18},
	pages={1--45},
	year={2017}
}

@article{Ruiz2020shopper,
	title={SHOPPER: a probabilistic model of consumer choice with substitutes and complements},
	author={Ruiz, F. J. R. and Athey, S. and Blei, D. M.},
	journal={The Annals of Applied Statistics},
	volume={14},
	number={1},
	year={2020},
	pages={1--27}
}

@article{heiss2021,
    author = {Heiss, F. and McFadden, D. and Winter, J. and Wuppermann, A. and B. Zhou},
    title = {Inattention and switching costs as sources of inertia in Medicare Part D},
    journal = {American Economic Review} ,
    volume={111},
    number={3},
    pages={2737--2781},
    year = {2021}
}

@article{hortaccsu2017power,
  title={Power to choose? An analysis of consumer inertia in the residential electricity market},
  author={Horta{\c{c}}su, Ali and Madanizadeh, Seyed Ali and Puller, Steven L},
  journal={American Economic Journal: Economic Policy},
  volume={9},
  number={4},
  pages={192--226},
  year={2017}
}

@article{bronnenberg2012evolution,
  title={The evolution of brand preferences: Evidence from consumer migration},
  author={Bronnenberg, Bart J and Dub{\'e}, Jean-Pierre H and Gentzkow, Matthew},
  journal={American Economic Review},
  volume={102},
  number={6},
  pages={2472--2508},
  year={2012}
}

@article{mackay2024consumer,
  title={Consumer inertia and market power},
  author={MacKay, Alexander and Remer, Marc},
  journal={Available at SSRN 3380390},
  year={2024}
}

@article{guadagni1983logit,
  title={A logit model of brand choice calibrated on scanner data},
  author={Guadagni, Peter M. and Little, John D. C.},
  journal={Marketing Science},
  volume={2},
  number={3},
  pages={203--238},
  year={1983}
}

@article{keane1997heterogeneity,
  title={Modeling heterogeneity and state dependence in consumer choice behavior},
  author={Keane, Michael P.},
  journal={Journal of Business \& Economic Statistics},
  volume={15},
  number={3},
  pages={310--327},
  year={1997}
}

@book{train2009discrete,
  title={Discrete Choice Methods with Simulation},
  author={Train, Kenneth E.},
  edition={2},
  publisher={Cambridge University Press},
  address={Cambridge},
  year={2009}
}

@incollection{mcfadden1974conditional,
  author    = {McFadden, Daniel},
  title     = {Conditional logit analysis of qualitative choice behavior},
  booktitle = {Frontiers in Econometrics},
  editor    = {Zarembka, Paul},
  publisher = {Academic Press},
  address   = {New York},
  year      = {1974},
  pages     = {105--142}
}

@article{berry1995automobile,
  author  = {Berry, Steven and Levinsohn, James and Pakes, Ariel},
  title   = {Automobile Prices in Market Equilibrium},
  journal = {Econometrica},
  volume  = {63},
  number  = {4},
  pages   = {841--890},
  year    = {1995}
}

@article{hoffman2013stochastic,
  author  = {Hoffman, Matthew D. and Blei, David M. and Wang, Chong and Paisley, John},
  title   = {Stochastic Variational Inference},
  journal = {Journal of Machine Learning Research},
  volume  = {14},
  pages   = {1303--1347},
  year    = {2013}
}

@inproceedings{ranganath2014blackbox,
  author    = {Ranganath, Rajesh and Gerrish, Sean and Blei, David M.},
  title     = {Black Box Variational Inference},
  booktitle = {Proceedings of the Seventeenth International Conference on Artificial Intelligence and Statistics (AISTATS)},
  year      = {2014}
}

@inproceedings{kingma2014autoencoding,
  author    = {Kingma, Diederik P. and Welling, Max},
  title     = {Auto-Encoding Variational Bayes},
  booktitle = {Proceedings of the International Conference on Learning Representations (ICLR)},
  year      = {2014}
}

@article{carpenter2017stan,
  title   = {Stan: A probabilistic programming language},
  author  = {Carpenter, Bob and Gelman, Andrew and Hoffman, Matthew D. and Lee, Daniel and Goodrich, Ben and Betancourt, Michael and Brubaker, Marcus and Guo, Jiqiang and Li, Peter and Riddell, Allen},
  journal = {Journal of Statistical Software},
  volume  = {76},
  number  = {1},
  pages   = {1--32},
  year    = {2017}
}

@article{dube2010state,
  title   = {State dependence and alternative explanations for consumer inertia},
  author  = {Dub{\'e}, Jean-Pierre and Hitsch, G\"unter J. and Rossi, Peter E.},
  journal = {The RAND Journal of Economics},
  volume  = {41},
  number  = {3},
  pages   = {417--445},
  year    = {2010}
}

@incollection{heckman1981,
  author    = {Heckman, James J.},
  title     = {The incidental parameters problem and the problem of initial conditions in estimating a discrete time-discrete data stochastic process},
  booktitle = {Structural Analysis of Discrete Data with Econometric Applications},
  editor    = {Manski, Charles F. and McFadden, Daniel},
  publisher = {MIT Press},
  address   = {Cambridge, MA},
  year      = {1981},
  pages     = {179--195}
}

@article{nevo2000mergers,
  title   = {Mergers with differentiated products: The case of the ready-to-eat cereal industry},
  author  = {Nevo, Aviv},
  journal = {The RAND Journal of Economics},
  volume  = {31},
  number  = {3},
  pages   = {395--421},
  year    = {2000}
}

@article{koren2009matrix,
  title   = {Matrix Factorization Techniques for Recommender Systems},
  author  = {Koren, Yehuda and Bell, Robert and Volinsky, Chris},
  journal = {Computer},
  volume  = {42},
  number  = {8},
  pages   = {30--37},
  year    = {2009}
}

@inproceedings{salakhutdinov2008probabilistic,
  title     = {Probabilistic Matrix Factorization},
  author    = {Salakhutdinov, Ruslan and Mnih, Andriy},
  booktitle = {Advances in Neural Information Processing Systems},
  year      = {2008}
}

@article{anderson1956statistical,
  title   = {Statistical inference in factor analysis},
  author  = {Anderson, T. W. and Rubin, Herman},
  journal = {Proceedings of the Third Berkeley Symposium on Mathematical Statistics and Probability},
  volume  = {5},
  pages   = {111--150},
  year    = {1956}
}

@article{lopes2004factor,
  title   = {Bayesian factor analysis and density estimation using a class of factor models},
  author  = {Lopes, Hedibert F. and West, Mike},
  journal = {Biometrika},
  volume  = {91},
  number  = {4},
  pages   = {719--733},
  year    = {2004}
}

@article{blei2017variational,
  author  = {Blei, David M. and Kucukelbir, Alp and McAuliffe, Jon D.},
  title   = {Variational Inference: A Review for Statisticians},
  journal = {Journal of the American Statistical Association},
  volume  = {112},
  number  = {518},
  pages   = {859--877},
  year    = {2017}
}

@article{dubois2022use,
  title={The use of scanner data for economics research},
  author={Dubois, Pierre and Griffith, Rachel and O'Connell, Martin},
  journal={Annual Review of Economics},
  volume={14},
  number={1},
  pages={723--745},
  year={2022},
  publisher={Annual Reviews}
}

@article{ackerberg2001yogurt,
  title={Empirically distinguishing informative and prestige effects of advertising},
  author={Ackerberg, Daniel A},
  journal={RAND Journal of Economics},
  pages={316--333},
  year={2001}
}

@article{villas2007yogurt,
  title={Vertical relationships between manufacturers and retailers: Inference with limited data},
  author={Berto Villas-Boas, Sofia},
  journal={The Review of Economic Studies},
  volume={74},
  number={2},
  pages={625--652},
  year={2007}
}

@article{hristakeva2022yogurt,
  title={Vertical contracts with endogenous product selection: An empirical analysis of vendor allowance contracts},
  author={Hristakeva, Sylvia},
  journal={Journal of Political Economy},
  volume={130},
  number={12},
  pages={3202--3252},
  year={2022}
}

@techreport{backus2021cereal,
  title={Common ownership and competition in the ready-to-eat cereal industry},
  author={Backus, Matthew and Conlon, Christopher and Sinkinson, Michael},
  year={2021},
  institution={National Bureau of Economic Research}
}

@techreport{atalay2023scalable,
  title={Scalable demand and markups},
  author={Atalay, Enghin and Frost, Erika and Sorensen, Alan T and Sullivan, Christopher J and Zhu, Wanjia},
  year={2023},
  institution={National Bureau of Economic Research}
}

@article{dopper2025rising,
  title={Rising markups and the role of consumer preferences},
  author={D{\"o}pper, Hendrik and MacKay, Alexander and Miller, Nathan H and Stiebale, Joel},
  journal={Journal of Political Economy},
  volume={133},
  number={8},
  pages={2462--2499},
  year={2025}
}

@article{mcfadden2000mixed,
	title={Mixed MNL Models For Discrete Response},
	author={McFadden, D. and Train, K.},
	journal={Journal of Applied Econometrics},
	volume={15},
	pages={447--470},
	year={2000}
}

@misc{Betancourt2018conceptual,
	title={A Conceptual Introduction to Hamiltonian Monte Carlo}, 
	author={Betancourt, M.},
	year={2018},
	eprint={1701.02434},
	archivePrefix={arXiv},
	primaryClass={stat.ME}
}

@article{croissant2020mlogit,
    title = {Estimation of Random Utility Models in {R}: The {mlogit}
      Package},
    author = {Yves Croissant},
    journal = {Journal of Statistical Software},
    year = {2020},
    volume = {95},
    number = {11},
    pages = {1--41}
}

@article{kleibergen2003bayesian,
  title={Bayesian and classical approaches to instrumental variable regression},
  author={Kleibergen, Frank and Zivot, Eric},
  journal={Journal of Econometrics},
  volume={114},
  number={1},
  pages={29--72},
  year={2003}
}

\clearpage


\appendix
\renewcommand{\thesubsection}{\Alph{subsection}}
{\noindent\LARGE\textbf{Appendix}}
\subsection{Prior robustness analysis}\label{a: prior_check}
\begin{figure} [h]

    \centering
    \includegraphics[width=1\textwidth]{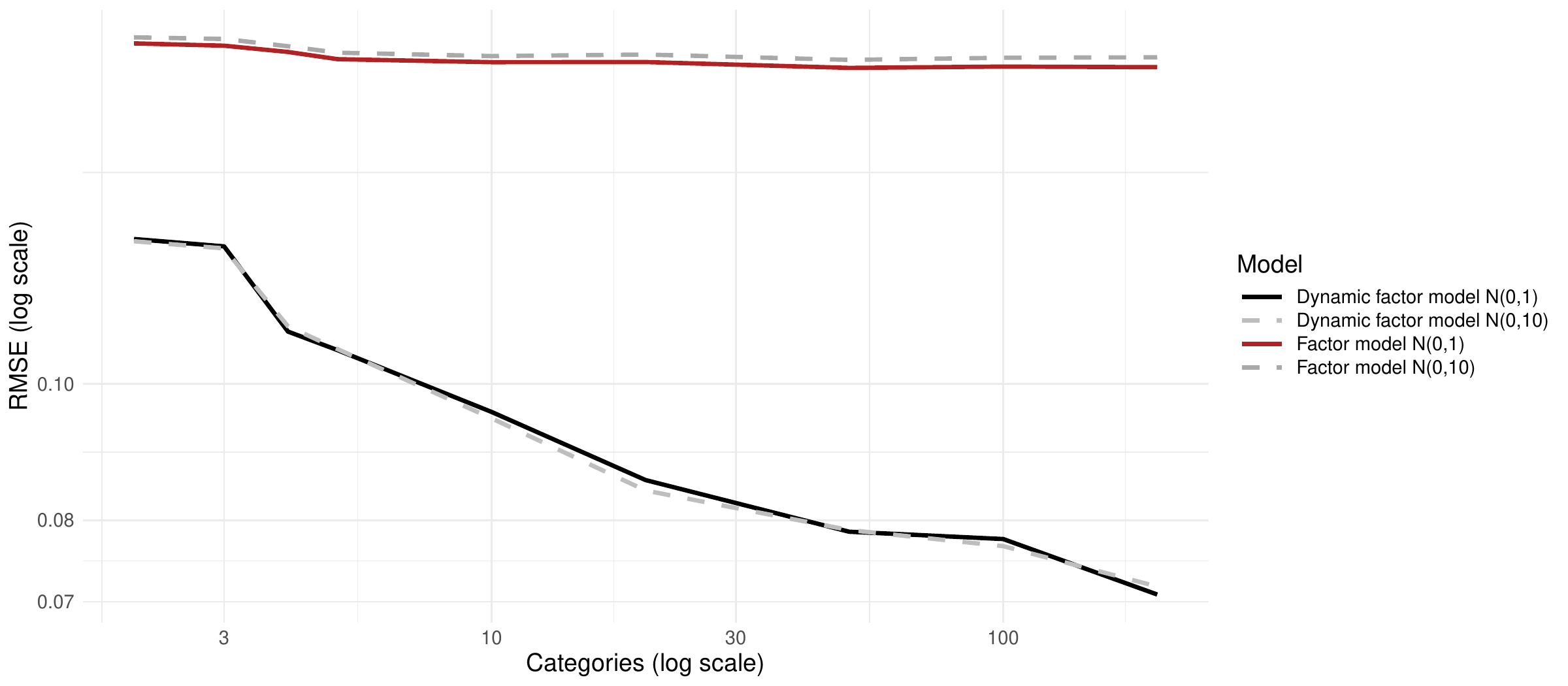}
    \caption{Simulation results for standard normal priors and less informative $\mathcal{N}(0,10)$ priors.}
    \label{fig: prior_check}
\end{figure}

\end{document}